# Incorporating Multi-Agent Systems Technology in Power and Energy Systems of Bangladesh: A Feasibility Study

Syed Redwan Md Hassan[1*], Nazmul Hasan[2*], Mohammad Ali Siddique[3], K.M Solaiman Fahim[4], Rummana Rahman[5], Lamia Iftekhar[6]

*Department of Electrical and Computer Engineering, North South University, Dhaka, Bangladesh*

[1]redwan.hassan@northsouth.edu , [2]nazmul.hasan05@northsouth.edu, [3]mohammad.siddique01@northsouth.edu, [4]solaiman.fahim@northsouth.edu, [5] rummana.rahman@northsouth.edu, [6]lamia.iftekhar@northsouth.edu

*Abstract—* The power sector of Bangladesh is presently experiencing essential changes as demand for power services is increasing with rising population and economic development. With a gradual shift from a rigidly centralized structure to a more decentralized and fluid setup, fundamentally because of the enormous advancement of distributed renewable energy sources, the future power system of the nation requires new control strategies to work efficiently and sustainably in the face of evolving conditions and constraints. Multi-Agent Systems (MAS) technology has attributes that meet these prerequisites of modern power systems and has been shown to be effective in dealing with its distributed and complex nature. This is a literature-based feasibility study to explore whether MAS technology is suited to be applied in the context of Bangladesh. For this preliminary paper, we look at the topic from a holistic perspective and conduct a meta-review to curate common applications of Multi-Agent System-based concepts, tools and algorithms on the power and energy sector. We also identify the top challenges of this domain in Bangladesh and connect the potential MAS-based solutions to address each challenge. Our qualitative assessment is motivated to provide a starting point for local researchers eager to experiment with MAS technology for application in Bangladesh.

I. INTRODUCTION

Bangladesh is rated as one of the fastest developing economies of South Asia, with 64% of the total population living in rural regions [1]. By 2020, Bangladesh ranked as 8th among the world's most populated countries. Although the country has made significant progress in terms of widening electricity coverage over the past few years, around 25 million people are still without access to electricity [2]. Electrification and having opportunities to use it effectively, are crucial for improving standard of living and maintaining the economic growth of the country in the long run. Thus, it is of utmost importance to actively explore and utilize the best technology and solutions to address existing challenges in the country's power and energy sector and ensure the continued sustainable development of the nation.

Currently, the majority of the electricity in the country stem from a few centralized generation systems [3]. However, centralized power systems are usually inflexible, incur high distribution costs, suffer from efficiency issues, often fail to reach out to remotest areas, and have major environmental impacts [5]. Moreover, such systems are vulnerable to single point of failure attacks, which may devastatingly impact large portions of the population and pose a national threat. It is thus imperative for us to start evolving towards active distribution networks and innovations in electricity transmission, including advanced future network technologies and distributed intelligent systems [3,4,5].

For the last several decades, decentralized and distributed systems and the large complex networked systems they are modelled as, have been researched upon extensively along with the latest intelligent technologies. For decades, intelligent technologies have been used to address a wide range of power system problems, using a general inclination of smart techniques, such as, multiagent systems, knowledge discovery, machine learning, deep learning, cyber-physical systems, cognitive informatics etc. [6]. In this work, we focus on exploring Multi-Agent Systems (MAS) technology and study its usage in the power and energy applications. Multi-Agent Systems concepts and methods, with their inbuilt flexibility, scalability, resilience, modularity and various other attractive characteristics, promise to be effective in managing the advanced challenges of modern power systems and enormous and multifaceted aspects of distributed energy setups [5,7,8].

It is intriguing to see whether the immense benefits of using Multi-Agent Systems technology applied to the power and energy sectors of developed nations can be translated to Bangladesh's power system. With that motivation, we explore whether MAS technology is a good fit to address the power and energy challenges of the country. This paper is a literature-review-based preliminary feasibility study.

The rest of the paper is organized as follows. Section II provides a brief overview of the extant literature to establish the gap in existing works which we are attempting to fill. Section III (Methodology) explains how we approached this study. Section IV presents a discussion of MAS technology in power and energy systems and curates and organizes concepts and terminologies. Section V identifies the major energy problems in Bangladesh and directs the reader to how they can be addressed using MAS technology. Concluding remarks are provided in Section VI.

II. PAST WORK

We went through an extensive review of the existing works of Multi-Agent System applications in the power domain and the recent works based on Bangladesh power systems. In order to situate our work in the literature, we can divide the past works in two categories- potential application of MAS in power and MAS scopes in the power and energy sector of Bangladesh.

---

\* These two authors contributed equally to this paper.

## A. Applications of MAS in Power

Multi-Agent Systems (MAS) technology is a vast technical field, and it has proven to be a promising area in the field of power systems. In recent years, MAS has received tremendous attention for its range of applications, such as system monitoring & diagnostics, power system restoration, market simulation, network control and automation as well as planning and design (see [7], [8], [10], [12] and the references therein). Moreover, as renewable energy-based smart grid and microgrid systems rise in popularity, MAS technology has been establishing itself as the paradigm of choice for modelling, analysis, control and optimization of such systems [9,11]. It is easily observed that comprehensive review works have been carried out every few years on the application of MAS in Power, and it is important to continue this trend, as the power sector is evolving rapidly and a holistic perspective is needed periodically to evaluate where we stand.

## B. MAS scopes in Power Sector of Bangladesh

It Bangladesh, although extensive MAS-based projects have not been undertaken, several experimental works in the nation's power and energy context have been applied employing certain aspects of MAS technology.

A popular topic of focus has been microgrid systems which have been found to be befitting for the rural, hilly, coastal, and remote areas of Bangladesh, where grid connectivity is very costly [13-17]. As microgrids are usually networked systems in nature, their control and modelling often fall within the MAS domain. Other works on agent-based technology in Bangladesh includes: managing load shedding using Ant colony optimization (ACO) algorithm [18], locating fault points using wireless sensor networks [19] and market pricing optimization [17].

MAS has shown an incredible potential in the global power system towards its transformation journey to a smart grid. Whereas, MAS can be a promising answer to many of the sustaining issues in the management of Bangladesh grid network, Bangladesh power literatures still lack in the thorough analysis and system modelling implementing MAS.

This paper contributes in the exploration of the literature on local power and energy systems to curate the top challenges in Bangladesh power systems at the moment. We try to establish a tangible connection between each of the local challenges and the relevant MAS –derived solution.

## III. METHODOLOGY

We have penned this preliminary survey paper through comprehensive analysis of secondary sources. Examining over journal, conference and book chapter publications, we grasped the idea of the status quo on MAS applications in the power sector globally. Next, we curated and amalgamated concepts and categorizations from several survey papers and customized them for this paper to set a premise for our main contribution (Table 1 and its connection to Section V). Choosing the demarcations of 'tools', 'computational methods or algorithms' and 'models or architecture', we recorded all items that we came across through the perusal of the papers and filed them under the relevant category. This enabled us to map such solution items with specific challenges in the power sector of Bangladesh (which were also curated simultaneously during the paper perusal) and provide a qualitative assessment on the potential use of MAS in the power sector of Bangladesh. Our secondary sources were collected from reputed repositories such as IEEE Xplore and the majority of the sources were published within the last ten years.

## IV. MAS TECHNOLOGY IN POWER AND ENERGY SYSTEMS

In this section we give a brief overview of MAS technology which will help us set the premise for solutions for specific issues in Bangladesh power systems in the next section.

## A. What is Multi-Agent Systems Technology?

Multi-Agent Systems are a setup of several agents (objects, nodes, people, robots, machines etc.), who can communicate among each other, to work towards solving a task collectively. Compared to a generic software or hardware unit, an agent is considered to have the following qualities: it can take action on its own (autonomous), it can cooperate with others to move towards a goal, compromising its immediate advantage in order to achieve a greater good, if needed (social), it can handle the environment (reactive), it is fast enough to work in tandem with others in the face of continuously changing situations (responsive) and it is always deciding and taking action towards a goal (proactive) [9, 20].

The study of Multi-Agents Systems has given rise to numerous models, concepts, frameworks, architectures, simulation tools, computational methods and algorithms - all of which we are terming collectively as MAS technology. Multi-Agent Systems inherently bring several advantages with them, the most significant among them being flexibility and resiliency. Such systems are usually modular in nature making it flexible when it comes to adding or modifying new aspects or building upon existing infrastructure. Unlike a centralized system, multi-agent system's inherent distributed nature implies that failure at one part of the system will not jeopardize the entire system making the Multi-Agent systems quite resilient.

Given these characteristics, it is recommended the multi-agent systems technology should be considered for complex systems which consists of a large number of entities communicating with each other, and modelling the overall system behavior in explicit terms would be highly difficult. Moreover, this technology is optimal for cases where local data is easily available and reliable, and where new modules and functionalities need to be added over time.

Power and energy systems present an excellent application paradigm for multiagent systems technology, as modern power and energy setups and requirements demonstrate all of the characteristics above.

## B. MAS in Power System

The idea of using MAS technology in power and energy systems have been explored for quite some time (as explained in Section II Part A), and a standardization attempt was undertaken in 2007. A Working Group on Multi-Agent Systems was formed in IEEE Power and Energy Society (IEEE PES) around that time, which generated a comprehensive direction for utilizing MAS technology in this domain [8][21]. The Working Group provided various top-down concepts and categorizations that helped later review papers place themselves in the literature. For our work, we are heavily inspired by a time-scale-based categorization of one such review paper. Moradi et. al [7] provide seven paradigms of power systems (see Column 1 of Table 1),

namely- protection, control, monitoring, network operation, network management, electricity market and planning; arranged according to the associated time horizon (starting in the scale of milliseconds for protection and ending at the scale of a decade or more for planning).

Modern power and energy systems are inherently complex networks that require strong communication among its constituents. Multiagent systems have been shown to be effective with its modeling and control tools for such networks in other domains - and the power sector is not to be left behind. Advances in technology, deregulation and competition, increased awareness in environmental concerns and security concerns have been catalysts in causing a shift in modern power systems towards a more distributed nature. With that, came in a flurry of experiments of applying MAS oriented tools, standards and techniques.

We have studied numerous review and original papers on multi agent systems technology published in the last fifteen years and extracted frequently occurring MAS-based models, tools and computational methods (Table 1). These items are mapped against their common power and energy application and further mapped against the seven paradigms provided by [7]. It can be observed from the table that various pre-existing MAS focused algorithms have been comfortably adapted for power engineering applications. On the other hand, new models and architectures have been designed from scratch which applies the advantages of MAS concepts in power systems.

TABLE I. EXISTING MAS TECHNOLOGY BASED SOLUTIONS USED FOR VARIOUS POWER AND ENERGY APPLICATIONS

| Paradigm | PES Application | MAS Technology (Model/Architecture, *Tool*, Algorithm) |
|---|---|---|
| PLANNING | Generation expansion planning | Dynamic Programming [22] [23] |
| PLANNING | Utility automation planning | *FIPA-ACL (FIPA compliant Agent Communication Language)* [24] |
| PLANNING | Planning and designing of hybrid microgrids | *HOMER (Hybrid Optimization of Multiple Energy Resources)* [13][15][25][26], *MATLAB/Simulink* [14][15] |
| ELECTRICITY MARKET | Improve earning from generation | Agent-based Modelling of Electricity Systems (AMES) [27] |
| ELECTRICITY MARKET | Optimizing bidding (Electricity market negotiation) | Dynamic Game Theory [28], *Multiagent System Simulator for Competitive Electricity Market (MASCEM)* [29], Colored Petri-Nets [30], Transactive Control Mechanism [31, 32, 33] |
| NETWORK MANAGEMENT | Demand management, congestion management | *PowerMatcher* [34-35], *ASPECS (a simulator)* [36], *PROLOG*[37], *JADE* [10] |
| NETWORK MANAGEMENT | Microgrid management | Multi-Agent Based Distributed Energy System (MAS-DES)[38], *JADE* [39] |
| NETWORK OPERATIONS | Restoration | Multi-Agent Immune Algorithm (MIA) [43], Colored Petri-Nets [30], IntelliTEAM II Automatic Restoration System [44], INTEGRAL [45] |
| NETWORK OPERATIONS | Substation automation | *SCADA* [46], Autonomous Regional Active Network Management System (AuRA-NMS) [47 and its references] |
| NETWORK OPERATIONS | Load Shedding | Particle swarm optimization [40][41], Ant Colony Optimization [18], Strategic Power Infrastructure Defense (SPID) [42] |
| MONITORING AND DIAGNOSTICS | Network fault diagnosis (Network monitoring) | ARCHON (Architecture for Cooperative Heterogeneous On-line Systems) [49], PEDA (Protection Engineering Diagnostics Agents) [50], *SCADA* [50] |
| MONITORING AND DIAGNOSTICS | Plant Item fault diagnosis (condition monitoring) | Condition Monitoring Multi-Agent System (COMMAS) [51 and references therein], COMMAS + PEDA [52] |
| MONITORING AND DIAGNOSTICS | Load flow, power flow, nodal analysis | *NEPLAN* [53] |
| CONTROL | Generic network control and optimization | *MATLAB/ Simulink*, *OPAL-RT* [48], Autonomous Regional Active Network Management System (AuRA-NMS) [54] |
| CONTROL | Optimizing voltage profiles in transformers | Fuzzy agent algorithms [55] |
| CONTROL | Load frequency control | Genetic Algorithm [56] |
| CONTROL | Smart grid control | Multi-Agent System Service-Oriented Architecture MAS-SOA [57], *MACSimJx* [58], IP-based Multi Micro-Grid Control System (IP MAS MMCS) [59] |
| PROTECTION | Protection coordination in smart grid | MAS-ProteC [60], Microprocessor Based Protective Relays (MPR)[61], JADE[62] |
| PROTECTION | Wide-area backup protection | Colored Petri-Nets [30] |
| PROTECTION | Preventing cascade tripping | Strategic Power Infrastructure Defence System (SPID) [52] |

## V. MAS Possibilities for Bangladesh Power System (BPS)

The existing power system of Bangladesh is very complex and quite aged; with 106 power stations connected to the grid of the country, and sustaining many shortcomings with respect to the continuously rising demand for electrical energy. There are several major challenges in the current system, which we tried to enlist here (in *A*) in order to map with feasible MAS solutions (in *B*).

### A. Major Challenges in Current BPS

The power generation sector of Bangladesh still uses traditional approaches such as, single cycle method and manual processes in generator installation to the infinite bus, incurring an average of 15-17% efficiency loss and risks of black-outs [63, 64]. The current transmission system lacks a sophisticated control system, automatic restoration system and regulated reserves. These result in the inability to manage the large variation of energy demand (season dependency), maintain current flow consistency, frequency stability, reliable supply control and power quality with the rapid expansion of the network [65, 10].

The biggest obstacles towards proper and frequent analysis of the grid's health are the absence of data from the required components and a bi-directional communication system [66, 67]. Additionally, cybersecurity is becoming a growing concern with the upgrading and inclusion of high-rate data flow in the grid's SCADA network [68]. Above all these issues, the integration of sources (renewable, nuclear, microgrids) to the power system with a proper coordination is proving to be difficult due to the lack of a reliable and secured internal communication network [69].

If we summarize the present status, the major challenges of the Bangladesh Power System can be listed as below-
- Optimized load shedding
- Frequency control
- Restoration system
- Maintaining power quality
- Integration of sources to grid
- Lack of advanced communication infrastructure

### B. MAS Solutions to the Existing Challenges of BPS

As we looked into the possible solutions for the existing issues in Bangladesh's power grid, we found several MAS architecture attributes that might help in addressing the high-priority problems. These findings are described as below-

*B.1 Load Shedding*

Load shedding mainly occurs when the frequency becomes less than 50Hz because of less power generation than the demand. In Table I, we have mentioned some MAS approaches, e.g., Particle Swarm Optimization, ACO and SPID, which have been proposed for load-shedding optimization. NEPLAN is a power system analysis software which can be used to monitor the load flow. The genetic algorithm can also be used to optimize the load frequency and prevent under-frequency load shedding [40, 41, 42, 18, 56].

*B.2 Frequency Control*

Unstable system frequency impedes the security and stability, causing severe energy loss and damage to the equipment. Sometimes, a blackout can occur as a result of the load frequency exceeding the grid frequency borderline. As a possible remedy, genetic algorithm based solutions can be exercised which have been known to provide an optimal frequency magnitude, maintaining real and reactive power flowing in each transmission line of a power system [56].

*B.3 Restoration system*

Fast and reliable system restoration is critical in case of any power cut, blackout, or line faults to maintain business operations in a fast-growing economy as in Bangladesh. MAS offers network operation tools such as SPID, IntelliTEAM II Automatic Restoration System, INTERGAL, MARS, AOPCPN, and MIA Algorithm, which can provide real-time sensing, failure analysis, vulnerability assessment, communication, and self-healing control features which can effectively address this issue [30, 43, 44, 45].

*B.4 Maintaining Power Quality:*

With the rapid expansion of the existing power network in parallel to the increasing electricity demand of Bangladesh, maintaining the power quality has become more challenging. Multi-agent System (MAS) technology such as AuRA-NMS can build a network management solution, in which any change of a complex power system can be automatically adapted in an intelligent manner. Rapid adjustment of line power flow can be controlled by a fuzzy agent algorithm with optimizing the voltage profiles in transformers [47, 54, 55].

*B.5 Integration of sources to grid*

MAS is developing as a new paradigm for controlling and managing grid integration. INTEGRAL and MAS-ProteC are powerful computational tools which can support the grid integration in Bangladesh. Dynamic Game Theory, MASCEM, and Transactive Control Mechanism are strong resources for electricity market design.

*B.6 Lack of advanced communication infrastructure*

For upgrading and overcoming the limitations of the communication infrastructure in Bangladesh, advanced tools are crucial. In power systems, MAS technology will provide better outputs using tools like ARCHON, PEDA, and Geographic Information Systems (Table I) [49, 52, 46, 68].

## VI. Conclusion

In this paper, we explored the application of Multi-Agent Systems (MAS) technology on Power and Energy Systems (PES). Taking a preliminary look at the status quo of Bangladesh's power and energy situation, we demonstrated that MAS technology has the potential to address each of the top challenges faced by the country's power sector at the moment. We provided a table with common MAS technologies for easy reference and mapping for future work experimenting with distributed power systems in all possible aspects, i.e., protection, control, monitoring, network operation and management, market and planning. This work is meant to be a preliminary study to obtain a bird's eye view of the intersection of MAS and PES and its possible application in Bangladesh. Future work would include more exhaustive and in-depth exploration as well as replication, adaptation and improvement of the most promising MAS technology from other countries on finding local and sustainable solutions.